# Harnessing functional segregation across brain rhythms as a means to detect EEG oscillatory multiplexing during music listening

**Dimitrios A. Adamos*[1,3], Nikolaos A. Laskaris[2,3], and Sifis Micheloyannis[4]**

*Asterisk indicates corresponding author*

[1]School of Music Studies, Aristotle University of Thessaloniki, 54124 Thessaloniki, Greece
[2]AIIA Lab, Department of Informatics, Aristotle University of Thessaloniki, 54124 Thessaloniki, Greece
[3]Neuroinformatics GRoup, Aristotle University of Thessaloniki, Greece - http://neuroinformatics.gr
[4] Medical Division (Laboratory L. Widén), University of Crete, 71409 Iraklion, Crete, Greece

Correspondence to:
Dr. Dimitrios A. Adamos
School of Music Studies, Faculty of Fine Arts,
Aristotle University of Thessaloniki,
GR-54124 Thessaloniki, GREECE
Tel: +30 2310 991839
Fax: +30 2310 991815
E-mail: dadam@mus.auth.gr, d.adamos@ieee.org

# Abstract

*Objective*. Music, being a multifaceted stimulus evolving at multiple timescales, modulates brain function in a manifold way that encompasses not only the distinct stages of auditory perception, but also higher cognitive processes like memory and appraisal. Network theory is apparently a promising approach to describe the functional reorganization of brain oscillatory dynamics during music listening. However, the music induced changes have so far been examined within the functional boundaries of isolated brain rhythms.

*Approach.* Using naturalistic music, we detected the functional segregation patterns associated with different cortical rhythms, as these were reflected in the surface EEG measurements. The emerged structure was compared across frequency bands to quantify the interplay among rhythms. It was also contrasted against the structure from the rest and noise listening conditions to reveal the specific components stemming from music listening. Our methodology includes an efficient graph-partitioning algorithm, which is further utilized for mining prototypical modular patterns, and a novel algorithmic procedure for identifying "switching nodes" (i.e. recording sites) that consistently change module during music listening.

*Main results.* Our results suggest the **multiplex** character of the music-induced functional reorganization and particularly indicate the dependence between the networks reconstructed from the $\delta$ and $\beta_H$ rhythms. This dependence is further justified within the framework of nested neural oscillations and fits perfectly within the context of recently introduced cortical entrainment to music.

*Significance.* Complying with the contemporary trends towards a multi-scale examination of the brain network organization, our approach specifies the form of neural coordination among rhythms during music listening. Considering its computational efficiency, and in conjunction with the flexibility of *in situ* electroencephalography, it may lead to novel assistive tools for real-life applications.

## Keywords

# Introduction

One of the most fascinating phenomena of real-world complex systems is the ability to demonstrate collective behavior from the synergy of their constituent compartments. During the last two decades, network-theory has successfully underpinned the cardinal role of *single-layer* intrinsic network interactions in the emergence and shaping of complex behavior; this contributed importantly to our understanding of biological, financial and various other social and technical systems [1,2].

More recently, network scientists have focused on the *multi-layer* characteristics of real-world systems, acknowledging the importance of encapsulating temporal and other context-related properties of the interactions themselves in describing coordinated function. Nowadays, scientists deal with the behavioral complexity of real-world systems by taking into consideration multiple coexisting interactions spanning different domains (indicatively see [3–6]). In a similar fashion, the human brain is also being characterized as a fundamentally multi-scaled complex network system [7–9]. The need for considering methods for handling such a multilayer organization has already been realized [10–13]. An obvious way is to associate each one of the brain rhythms with a distinct layer, then form an extended network by stacking the presumably communicating layers, and finally search for organization principles supporting cognition and behavior. In the case of **multiplex** networks, which model functional/effective connectivity, the inter-layer interactions are confined locally and concern only the replicas of each network site across layers [10,11].

Among the multiple aspects of human cognition, music perception has a fundamental role in human culture since the ancient ages. It is not surprising, that music perception and appraisal attracted the interest of scholars from diverse scientific fields, including psychology, philosophy and mathematics. Music cognition has been the subject of multiple studies during the last three decades (for details see [14,15] and remains in the spotlight of functional neuroimaging research (e.g. [16]). However, music constitutes a natural stimulus with rich temporal structure, recurrent patterns and semantics. Thus, among the neuroimaging modalities, the fast ones (i.e. the ones with high temporal resolution, like EEG/MEG) are expected to provide measurements useful for describing the impact of music on brain activity and tracking the induced cognitive processes.

While particular components of music (like pitch, rhythm and syntax) have been extensively studied [17], the use of naturalistic music stimuli has only recently attracted researchers' interest [18–21]. For most of the EEG studies employing music, the motivation stemmed from practical applications like multimedia implicit tagging (e.g. [22,23]) or from the power of music to bring subjects in distinct emotional states (e.g. [24,25]). There is, however, another important research direction which refers to understanding the effects of music on the neural synchrony and the functional organization of the cerebral networks. However, until now, all relevant research in the field has been confined within the boundaries of isolated brain rhythms (e.g. [26–32]).

Apparently, attempting a novel description of brain's functional (re)organization during music listening by adopting the methodological perspective of multiplex networking holds a lot of promise. One of the organization aspects, that is of great interest in adaptive complex systems, is the presence of functional modules, the so called *communities*[1] [33,34]. For functional cerebral networks in particular, community detection is currently considered an invaluable tool towards characterizing the multi-scale functional organization of the human brain [35]. Motivated by this fact, in this work, we developed an exploratory data analysis methodology that compares the community structure across layers in order to detect and delineate integrative organization trends in multi-layer networks reconstructed from experimental data.

The new algorithmic framework was applied to an experimental EEG dataset that contained brain activity recorded from 24 subjects, all exposed to the same music extract taken from the first part of *the Johann Sebastian Bach Violin Concerto No. 1, Allegro moderato in A minor (BWV 1041)*. The first movement of this concerto features as an excellent example of a fundamental musicological form known as the *ritornello form* [36,37]. The underlying musicological structure of this particular music extract remains stable[2] throughout its duration, demonstrating exemplary symmetry [38].

In a nutshell, data analysis proceeded in the following way. Multilayer functional networks were reconstructed from the signals, based on phase synchrony. A distinct single layer was associated with each brain rhythm and functional modules were detected for each layer independently. Inter-layer dependencies were sought by quantifying the similarity between the individual modular structures. Finally, we examined phase-amplitude coupling (PAC) as a possible mechanism for explaining the observed multiplexing. This last step was motivated by two emerging trends. Cross-frequency-interactions have been reported in studies of auditory processing based on sensor-level descriptions [23,39,40] and, very recently, cross-frequency coupling has been hypothesized to explain the multiplexed cortical organization throughout temporal scales [41].

---

[1] A community in graph-theoretic terms is defined as a group of mutually and strongly inter-connected nodes, which simultaneously are loosely connected to the rest of the network.

[2] We took advantage of this property when we selected artifact-free segments from the recordings.

The remainder of this paper is structured as follows. The experimental data and their preprocessing are first described. A short description of the adopted methodological ingredients is provided next and followed by the presentation of obtained results and a discussion of their interpretation and significance.

# Experimental data

## Experimental setup and EEG recordings

The present study concerned 24 right-handed non-musician volunteers (aged: 19–32 years, mean 24 years), who were Medical School students in the University of Crete. The study was approved by the local ethical committee and all participants signed an informed consent form after the procedures were explained to them. They had been asked to avoid alcohol (from the previous day) and caffeine (during the same day) consumption. The experiment included passive listening tasks during which the subjects had been instructed to remain still with eyes closed, since the purpose of the experiment was to search for EEG reflections of music listening that would emerge naturally, without the subject being engaged in any particular cognitive task.

Before starting the recording, each subject sat comfortably in an armchair and the loudness of the speakers was adjusted to a reasonable level. EEG was recorded continuously during three (3) conditions: a) Resting state with eyes closed (3 minutes) b) Listening to white noise (20 seconds) c) listening to part #1 of BWV 1041 - J.S. Bach Violin Concerto in A minor (20 seconds). During debriefing at the end of the recording session, each participant confirmed that he had experienced the two listening sections without any distraction.

The EEG was recorded from the following 29 electrodes according to the international 10/20 system: FP2, F4, F8, FC4, FT8, T4, TP8, C4, CP4, P4, PO8, O2, FP1, F3, F7, FC3, FT7, T3, TP7, C3, CP3, P3, PO7, O1, FZ, FCZ, CZ, PZ, OZ (see Fig. 1) and A1 + A2 as reference electrodes. The A1 and A2 electrodes were positioned on earlobes, far from the scalp electrodes (and far from muscles). The signals were amplified using a set of Contact Precision Instrument amplifiers, filtered online with a (0.03–200) Hz band pass and digitized at 500 Hz.

## Preprocessing.

EEG data were re-referenced using average-reference and filtered within [1-100 Hz] using a zero-phase band-pass filter (3rd order Butterworth filter). Artifact-free epochs of 8 s were selected via visual inspection for each subject and for each recording condition.

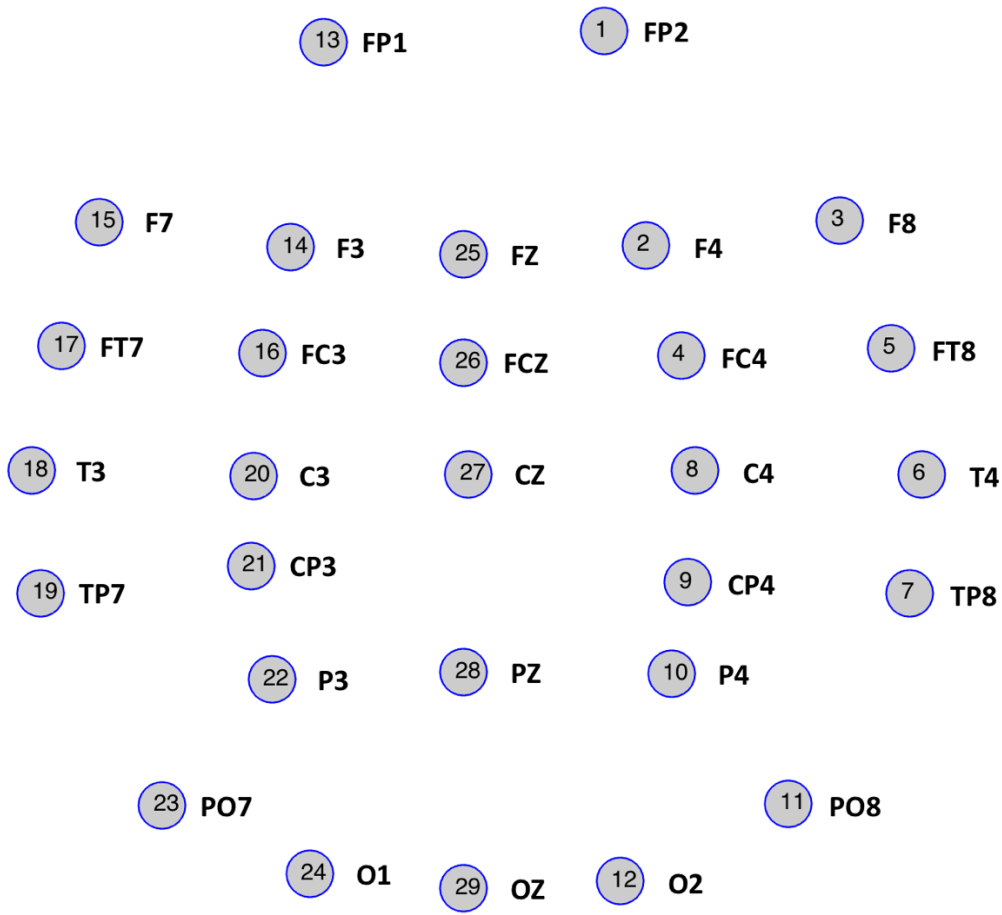

Figure 1: The positions and labels of EEG sensors used in our experiments

# Methods

## Outline

Below we provide a brief sketch of the methodological flow which is then accompanied with Figure 2 that serves as a coherent roadmap to the presented study. We commenced by adopting the standard practice to study the well-known brain rhythms independently. These rhythms were defined as follows. $\delta$ rhythm included brainwaves in (1–4) Hz band; $\theta$ rhythm brainwaves in (4–8) Hz; $\alpha$ rhythm in (8–13) Hz; *low-beta* or $\beta_L$ rhythm brainwaves in (13–20 Hz); *high-beta* or $\beta_H$ in (20–30 Hz); *low-gamma* or $\gamma_L$ in (30–45 Hz); *high-gamma* or $\gamma_H$ in (55–95 Hz). Specifically, the splitting of $\beta$ and $\gamma$ rhythms into their lower and higher components was motivated by our previous study on passive music listening [23] that revealed a particular functional role for the $\beta_H$ - $\gamma_L$ coupling. Local activation patterns associated with each rhythm were first derived and compared among recording conditions. Global patterns of coordinated activity were then derived, in the form of reconstructed functional networks, and the emerged rhythm-dependent connectivity patterns underwent modularity analysis. Subsequently the focus of analysis was turned onto examining the putative convergence among the functional organization-trends observed in the individual brain rhythms. With the emphasis put on functional segregation, and without the need for additional methodological steps, we smoothly transitioned to a multiplex network characterization, where the single-layers were the rhythm-depended functional networks and the goal was to detect inter-layer interactions that were specific to the music listening condition. Finally, we tested the hypothesis that the observed multiplexing trends could be attributed to PAC mechanism. A Matlab® source-code repository, including functions and scripts to demonstrate the essential algorithmic steps, is available through our groups website[3].

## EEG power spectrum descriptors

The power spectrum of the EEG signal, for each subject and at each sensor, was estimated in the 1 - 100 Hz frequency range based on FFT following Welch's method with a Hamming window. Signal power values for $\delta$ – $\gamma_H$ frequency bands were estimated by aggregating power spectral density (PSD) measurements within the bins corresponding to the 7 distinct brain rhythms, as defined above. In this way, three (3) spectral profiles were derived (per subject/sensor) indicating the frequency content of brain activation in the three recording conditions: Rest, Noise and Music. To provide a

---

[3] https://neuroinformatics.gr/research/publications

"first-order" estimation of the effect of (either the noise or) music, we first computed the median spectral profile for each recording condition and then formed the pattern of relative changes which was a 7-tuple of the form $(PSD_{listening}-PSD_{rest}/PSD_{rest})^{rhythm}$. The procedure was followed for each sensor independently and for contrasting both listening conditions to resting state. In addition, we employed Wilcoxon paired sign-rank test to assess the significance of listening induced changes in the signal power associated with each brain rhythm. To correct for the effect of multiple comparisons (7 rhythms × 29 sensors), the False Discovery Rate (FDR) method with α = 0.05 was used [42].

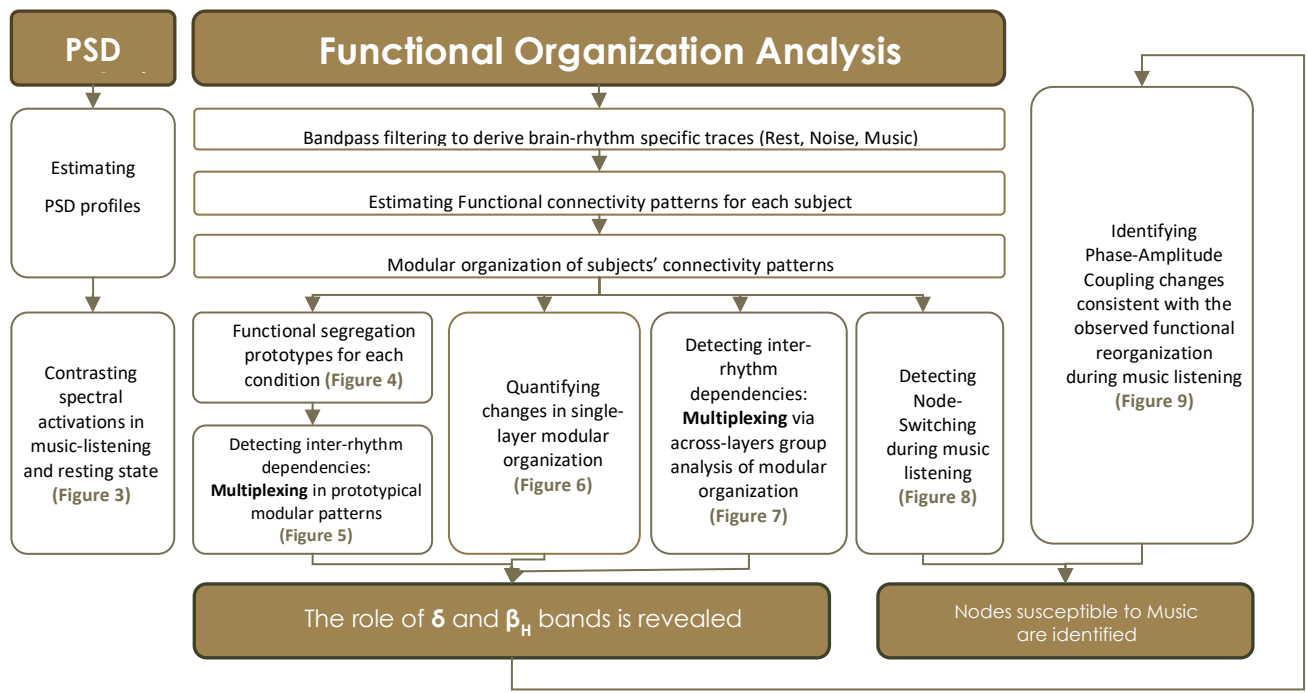

Figure 2: Methodological flowchart

## Functional Connectivity Patterns

Phase synchrony has been identified as a key mechanism that facilitates reciprocal interactions, large-scale integration and functional organization in the human brain[43]. Here we employed a popular estimator of phase synchrony to derive patterns of functional connectivity, embedded directly in sensor space, with the scope of detecting re-organization phenomena induced via music listening.

The Phase Locking Value (PLV) estimator [44] operates on a pair of band-limited signals and results to a value ranging from 0 to 1, with high values indicating high neural synchrony. It includes the estimation of instantaneous phase $\varphi_i(n)$ of each involved signal $x_i(n)$, $n=1...N$ (for instance by means of Hilbert Transform), the formation of phase differences $\Delta\varphi(n) = \varphi_{i_1}(n) - \varphi_{i_2}(n)$ and an integration step that finally outputs an index that quantifies the phase synchrony during the whole time interval under examination :

$$PLV^{x_{i_1},x_{i_2}}(n) = \left| \frac{1}{N} \sum_{n=1}^{N} \exp(i\,\Delta\varphi(n)) \right| \qquad (1)$$

In our study, PLV was applied after filtering the signals within a given frequency band that corresponded to each one of the examined brain rhythms using a 3rd order Butterworth filter (in zero-phase mode). Working in pairwise fashion, we quantified the synchrony for all possible combination of sensors. The derived quantities were tabulated in an [$N_{sensor} \times N_{sensor}$] matrix, in which an entry conveys the strength of the functional coupling between a particular pair of recording sites ($N_{sensor}=29$ is the number of recording sites). Each such matrix conveyed a functional connectivity pattern and had a natural graph-based representation, called hereafter *functional connectivity graph* (FCG), with nodes being the recording sites and edges representing the in-between links weighted by the tabulated value. For each subject and brain rhythm, three FCGs were created, one for each recording condition, and denoted respectively ($FCG^{rest}$, $FCG^{noise}$ and $FCG^{music}$).

## Functional Reorganization Analysis

### Functional Segregation Descriptors

The functional connectivity pattern associated with each FCG was analyzed, in terms of modular structure, using a flexible graph-theoretic, community-detection algorithm with a widely demonstrated ability of detecting cohesive graph components (e.g. [45]). The adopted algorithm is based on the notion of dominant sets [46] and operates on a given weighted adjacency graph (see Appendix 1). It has been previously proven very reliable in describing segregation in similar functional brain networks [47,48]. In the particular study it was used to provide a segregation profile for each FCG. The descriptor had the form of a 29-tuple c = [$c_1$,…, $c_{29}$], $c_i \in Z$ (e.g., c=[1 3 2 3 ••• 2 1 1 1 2]), with each integer representing the membership label of a particular sensor. These labels also denoted the relative-ordering of the detected modules (for instance the nodes with label 1 formed the most compact functional module), which was based on the *cohesiveness-index* (as described in Appendix 1).

### Functional Segregation Prototypes

To facilitate the comparison among recording conditions, we derived prototypical segregation profiles as representatives for each ensemble of functional connectivity patterns (24 FCGs per condition). To this end, the well-known technique of *Consensus-Clustering* [49,50] was applied to each ensemble using the code freely available from a recent study [51]. In short, based on an aggregation scheme that worked at the level of edges (and used all the 24 partitioned FCGs), we estimated the probability of any pair of nodes to occur in the same module. These empirical probabilities were used as weights in a new FCG, that was then fed to dominant-sets algorithm. The derived modular pattern (or equivalently graph partition) was representing the functional organization of a particular brain rhythm during in one of the recording conditions and can be thought of as a gross picture of how a single network layer was organized.

### Comparing Segregation Profiles: single-layer approach

To compare any two of the derived graph-partitions, the measure of Variation of Information (VI) [52] was employed. In general, VI operates on a pair of membership lists **c** and **c'** and quantifies the (dis)similarity between individual partitions (see Appendix 2). VI(**c**,**c'**) is a non-negative, symmetric distance which satisfies the triangular inequality and hence can be used to reliable express the "true" distances in partition space. It has already been successfully employed to compare community structure between (single-layered) functional brain networks [47,53].

In the present study, VI found a threefold use; we mention here the use, which is aligned with the notion of single-layer network analysis. VI was used to compare, for each subject and brain rhythm independently, the segregation profiles between rest condition and either music (or noise) listening.

### Comparing Segregation Profiles: multi-layer approach

VI was also utilized as a tool to detect important functional dependencies between the individual brain rhythms. It was employed in a multi-layer approach so as to unravel interrelationships between the organization patterns emerged in the corresponding reconstructed networks. This part of analysis was executed in two phases; one based on population summary and the other operated at a more detailed level. In the first phase, VI served for comparing the 7 distinct prototypical segregations (i.e. the rhythm-dependent graph partitions, see section 'functional segregation prototypes' above) emerged in a particular recording condition. A high similarity between two graph-partitions was considered as an indication of strong functional coupling between the brain rhythms. In the second phase, a more detailed comparison was performed. For each subject separately, the 7 segregation profiles were first compared against each other. A distribution of VI measurements was, then, formed for each rhythm-pair. Finally, the obtained measurements were statistically compared between recording conditions, by means of Wilcoxon rank-sum test. The FDR method was used for multiple comparison correction at $\alpha = 0.05$.

### Defining Switching Nodes

Since music induced changes in the modular pattern were of great importance, an algorithmic step was carefully designed in order to spot those nodes of the network that systematically changed their partnerships during music listening. The particular step was motivated by the recently introduced notion of *flexible nodes*, the existence of which had been demonstrated to be an important characteristic of functional brain dynamics underlying various complex cognitive processes [7]. With the scope of detecting the nodes that transitioned from their original functional community (during rest-state) to a different community (during music listening), and in order to overcome the problem of module correspondence between states (rest and music listening), we proceeded as follows (see also Fig.8). Working separately for each subject, and using the two partition profiles $c^{rest}$ and $c^{music}$, derived from the functional connectivity graphs $FCG^{rest}$ $FCG^{music}$ associated with a particular brain rhythm, we constructed two binary connectivity graphs that encapsulated the detected community structure. The corresponding binary adjacency matrices $\mathbf{A^{rest}}$ and $\mathbf{A^{music}}$ were sparse with ones only in the entries corresponded to pair of nodes sharing the same community label. By forming, elementwise, the absolute difference between them, we derived a binary matrix $\mathbf{A^{diff}} = \left|\mathbf{A^{music}} - \mathbf{A^{rest}}\right|$, in which all the pairs of nodes that changed partnership were listed. $\mathbf{A^{diff}}$ was compatible with the form of a [29×29] adjacency matrix representing a connectivity graph built over the 29 sensors. It was treated as such, and by deriving the node degrees in the induced graph we ended up with a node-dependent profile of switching propensity scores:

$$N_{SW \ [1 \times N_{sensors}]} = \sum_{i=1}^{N_{sensors}} \mathbf{A^{diff}}(i,j) \qquad (2)$$

These $N_{SW}$ scores were averaged across subjects to obtain an aggregate profile for each brain rhythm.

## Phase-Amplitude Coupling to assess interactions between brain rhythms

Since the assessment of cross-frequency coupling (CFC) has recently gained a lot of popularity in cognitive studies [54–56] and recently appeared as an alternative approach to study music-induced interactions between distinct brain rhythms [23,40], it was considered important to integrate it in the particular work. Among the various available CFC descriptors, phase-amplitude coupling (PAC) which relies on phase coherence and is the most commonly encountered descriptor [57,58], was selected. The employed PAC descriptor was implemented via estimating a phase-locking value (PLV), as described in Appendix 3. It resulted in a PAC-level lying between 0 and 1, with high values denoting a strong modulation of the amplitude of the brainwaves in the higher rhythm by the phase of the brainwaves in the lower rhythm. It was utilized with the scope of complementing the analysis of multiplexed segregation by suggesting PAC as explanatory mechanism for the observed dependencies among particular brain rhythms ($\delta \rightarrow \beta_H$). Working at the level of individual sensors, PAC-levels were estimated for all subjects and recording conditions. A distribution was then formed (from the 24 measurements) for each recording condition. The distribution of the rest condition was then statistically compared with the distribution of either the noise or the music listening condition, based on Wilcoxon paired sign-rank test. The FDR method was used for multiple comparison correction at $\alpha = 0.05$.

# Results

## Spectral changes induced by music

Our explorations started with the standard step of analyzing the spectral content of brain activations. Figure 3 contains the contrast between music listening and resting state, based on the aggregate spectral profiles derived from the ensemble of subjects, in the form of relative change for every brain rhythm. These changes are presented topographically together with an indication about their statistical significance (computed by comparing the corresponding distributions of PSD measurements). Among depicted results, one may observe the uniform decrease of power in the *$\alpha$-rhythm* and the increase in the *$\gamma_H$-rhythm power* localized in sensors over or nearby the auditory-cortex brain regions (e.g. FT7, TP7, FC3, FT8, T4), during music listening.

To justify the importance of the obtained results and test whether the observed changes could be actually attributed to music listening and not to the mere activation of the auditory pathway, we estimated the spectral changes during listening to noise as well (see Figure S1, in supplementary material). By comparing the trends in Fig.3 and Fig.S1, a similar pattern of spectral changes during both listening conditions becomes apparent.

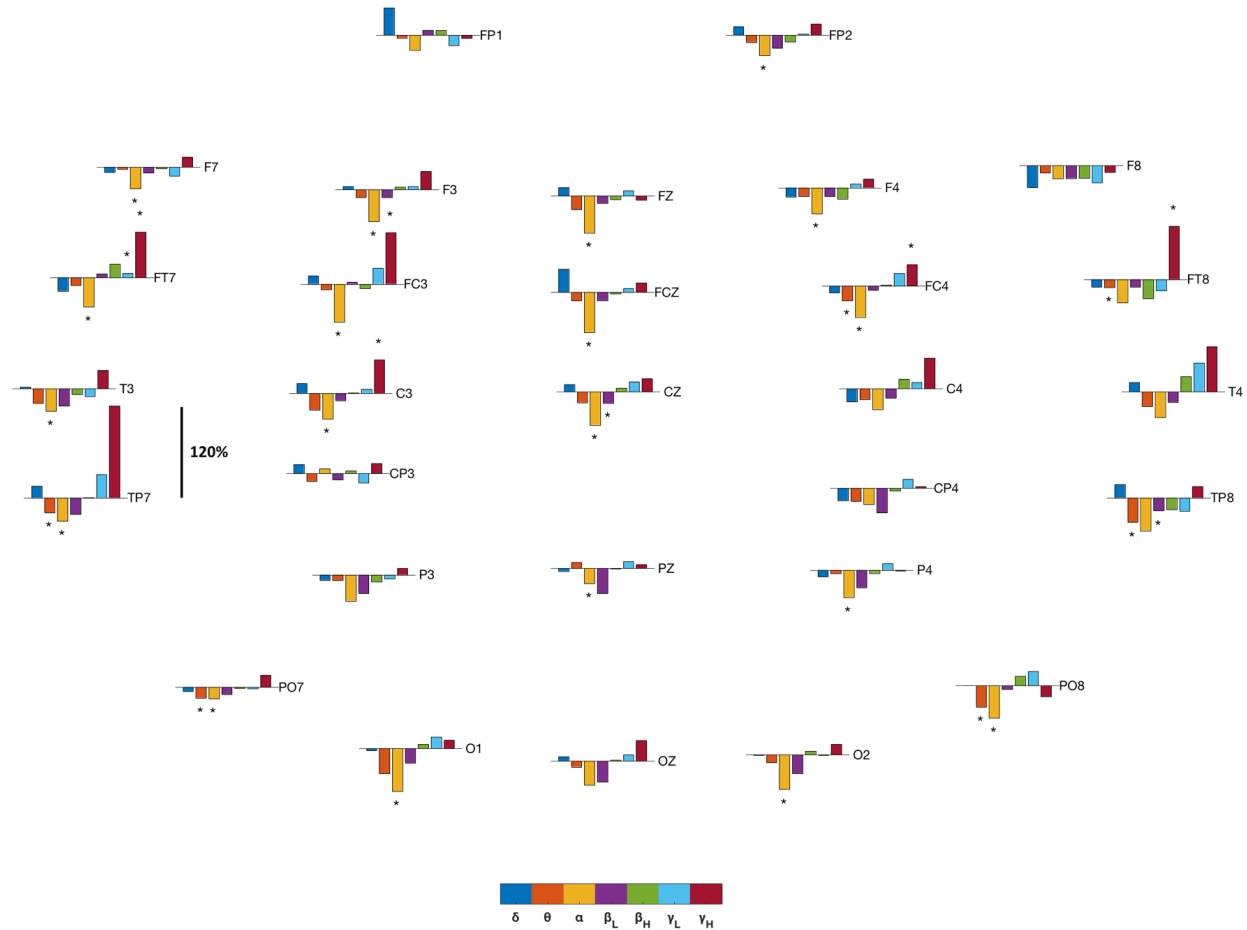

Figure 3: Contrasting spectral activations between Rest state and Music listening. At each sensor, the relative change in the spectral content of each brain rhythm is depicted. Statistically significant changes are denoted with '*' (p<0.01; FDR corrected).

## Functional Reorganization Analysis

**Probing music induced changes in inter-rhythm coupling based on aggregated descriptors of community structure**

Our quest for signatures of multiplex coordination started by expressing the similarity in terms of functional organization among the distinct networks reconstructed from the 7 brain rhythms and quantifying its change during music listening. To this end, we first followed a *Consensus Clustering* aggregation scheme (see methods) and derived prototypical patterns of modular organization for resting state and music listening. These are depicted in Figure 4 and represent a gross picture of the modular organization associated with each brain rhythm and recording condition (noise-listening related prototypes are also depicted for completeness). Presented in contrasting fashion, these prototypical segregation profiles enable the straightforward detection of self-organizing trends, like: a) the emergence of a strong functional module formed by occipital brain regions, consistently in the networks of all brain rhythms, and b) the widespread stable profile of the strongest community (nodes colored in blue) in *α-rhythm* which is tempting to associate it with the default-mode network. For the purpose of this study, of great interest was the music induced re-organization trend seen in $β_H$ brain rhythm, which was accompanied by a disruption of modular pattern during noise-listening (several nodes remain isolated from the detected communities).

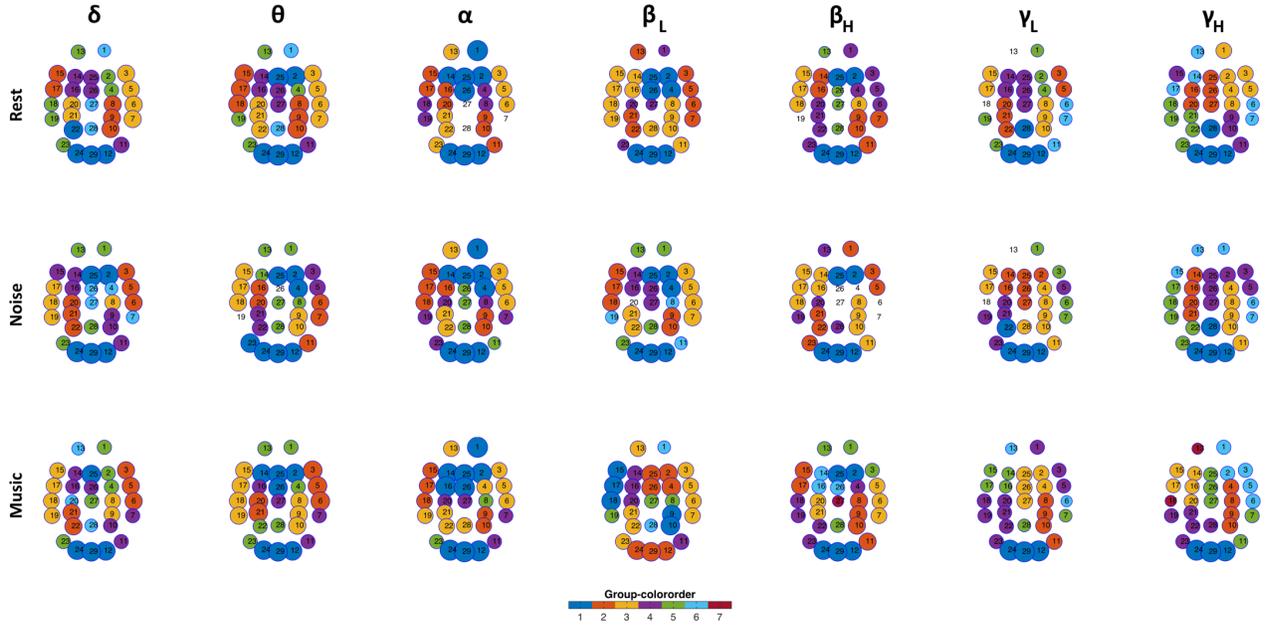

Figure 4: Prototypes of modular organization as a function of brain rhythm and recording condition. Nodes belonging to the same functional community share color and size. Color indicates the order of coherence among the formed groups (ranked from the most coherent to the least interconnected community). The size scales linearly with the *cohesiveness index* (see methods) and indicates the level of coherency of each community (using a common scale across all prototypes).

To provide a more refined picture, we employed a visualization strategy that exploited the descriptive power of VI metric to compare modular patterns quantitatively. The VI distances between every possible pair of prototypical segregations from a recording condition were computed and the corresponding measurements were tabulated accordingly in a [7×7] distance matrix $D_{VI}$. The distance-preserving technique of multidimensional scaling (MDS) was then applied, producing a mapping: $prototypical\_c_i \mapsto Y_i \in R^2$, i.e. $Y_{[7×2]}$=MDS($D_{VI}$). The mapping resulted in a two-dimensional scatter-plot that accommodated all pairwise comparisons. Two such scatter-plots, one for resting state and one for music listening, are shown in Fig.5. Each point corresponds to a band-specific modular prototype, and the inter-point distances reflect the differences in the functional organization among the corresponding brain rhythms. It becomes apparent that music induced changes in modular organization that made **δ** and **$β_H$** rhythms more similar in terms of functional segregation pattern. This observation provided some evidence about the functional dependence between the particular rhythms. It was considered suggestive of an increased inter-layer network covariation, and hence elevated multiplexing.

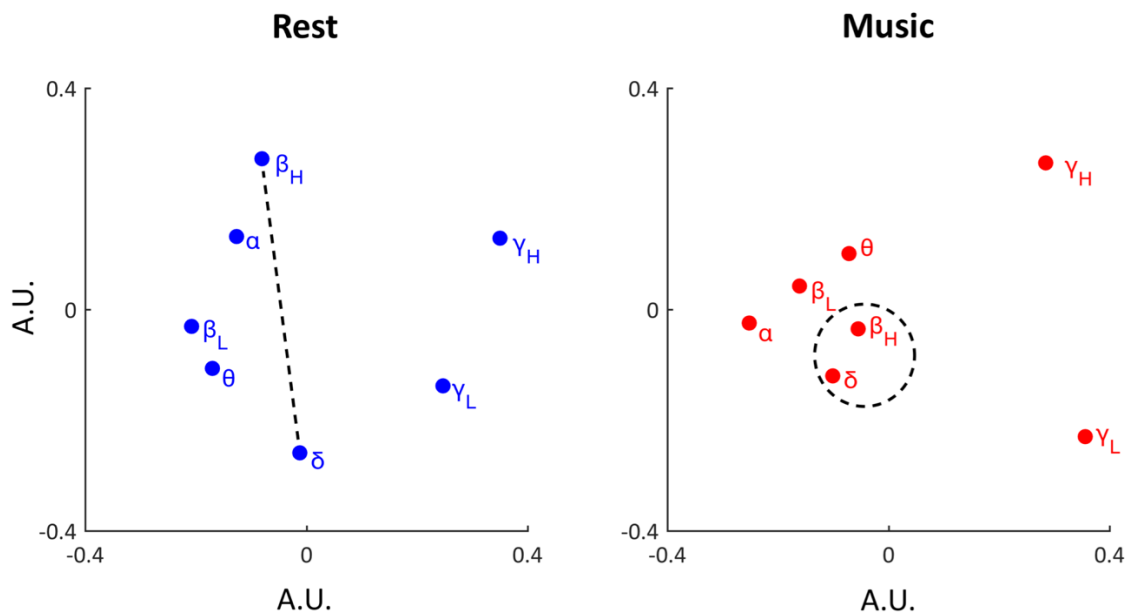

Figure 5: Visualizing the multiple pair-wise comparisons among the modular organization prototypes (shown in Fig. 4), based on the distance-preserving technique of MDS. In the derived scatter plots, each dot corresponds to a prototypical modular pattern associated with a brain rhythm and inter-dot distances reflect the corresponding deviations in terms of VI distance. All axes have been scaled equally and assigned the 'A.U.' label (in MDS-embeddings, the axes do not carry any particular physical meaning since the scope of MDS technique is to turn relational data to geometrical relationships). For each recording condition, a distinct scatterplot has been derived. Both scatterplots have been scaled so as to make inter-point distances comparable between these two plots.

**Quantifying within-rhythm changes in the modular reorganization during music listening**

To investigate deeper, the effect of music on functional re-organization, we returned to the individual modular patterns (24 segregation profiles per frequency band), and for each rhythm independently estimated the VI-distance between resting state and either music or noise listening. These 24 distances were used to form a distribution indicating the extend, in which a listening-condition altered the "baseline" modular pattern associated with the brain at rest. All these distributions are compared, based on whisker plots, in Figure 6. It is evident that music perturbs mostly the segregation in **δ** and **β$_H$** rhythms. Noise, on the other hand, appears to affect the organization in the **δ** (though in less extend) but not in **β$_H$** rhythm.

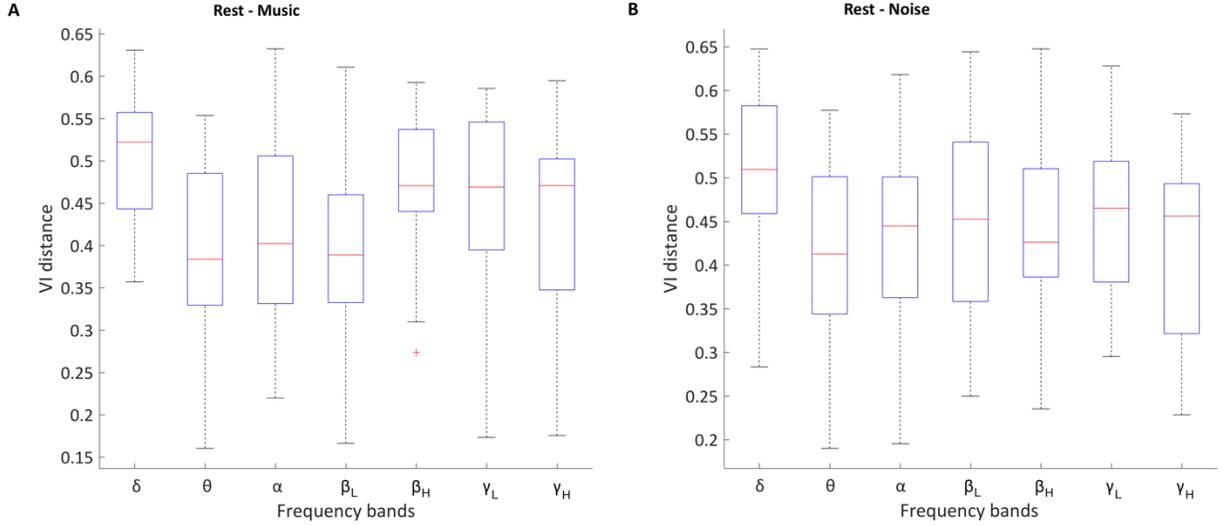

Figure 6. Quantifying the extend of alterations in modular organization by means of VI distance. Whisker plots are used to indicate the distribution, across all 24 subjects, of distances between community structure at rest and community structure during listening to either music (left) or noise (right).

**Detecting music induced changes in inter-rhythm coupling based on individual descriptors of community structure**

For a more thorough investigation of the inter-layer correlations regarding the emergence of communities, we examined the similarities among the individual (i.e. subject-specific) modular patterns in the following way. For each subject and recording condition separately, the 7 segregation profiles were initially compared against each other by means of VI-metric, resulting in a set of (7×6)/2=21 pairwise distances (per subject and condition). By collecting these measurements across subjects, a distribution of VI-measurements was then formed for each pair of rhythms and recording condition. For each pair, the distribution of VI measurements from resting state was compared with the distribution from a listening condition and the between-rhythm interactions corresponding to significant differences were detected (p<0.05; FDR corrected). Finally, the corresponding median values were estimated, and used to express the relative increase of inter-rhythm coupling with the following formula, where we took into consideration that VI, being a "distance", actually measures the dissimilarity between any two rhythms:

$$\text{Rel}_{\text{increace}}(\text{fb}_1, \text{fb}_2) = \frac{\text{median}\{\text{VI}(\text{fb}_1, \text{fb}_2)^{\text{rest}}\} - \text{median}\{\text{VI}(\text{fb}_1, \text{fb}_2)^{\text{listening}}\}}{\text{median}\{\text{VI}(\text{fb}_1, \text{fb}_2)^{\text{rest}}\}} \quad (3)$$

with $\text{fb}_{1,2} = \{\delta, \theta, \alpha, \beta_L, \beta_H, \gamma_L, \gamma_H\}$

Figure 7 includes these measurements in the form of a connectivity pattern that reflects the inter-layer increases/decreases in functional dependence for both listening conditions. Width and color encode complementary information about the multiplexing during music (Fig.7A) and noise listening (Fig.7b), which has been expressed as deviation from the ''baseline'' condition. The most notable observation is the increase in the functional coupling between **δ** and **β$_H$** rhythms, during music listening only. This figure provides ample evidence for an enhanced **δ-β$_H$ multiplexing**, as a result of music perception, that builds as a coherency in functional segregation.

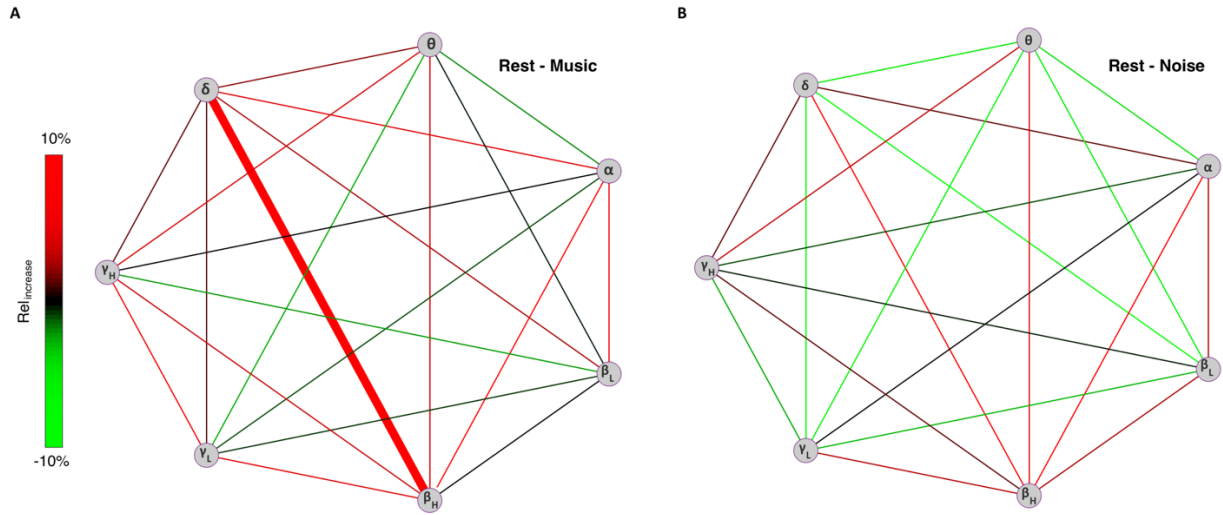

Figure 7. Detecting inter-rhythm coupling by measuring the VI distance between (the community structure of) every pair of brain rhythms. Measurements have been aggregated across subjects, for each recording condition separately, and then used to express the group-based relative changes. Left/Right includes a ''differential'' graph encapsulating the most important changes in between-rhythm coupling - hence **Multiplexing** - when listening to music/noise. Important changes (based on Wilcoxon rank-sum test; p<0.05, FDR-corrected) in between rhythm coupling are indicated with a thick-width line. Color indicates both the strength of normalized change (percentage) and the direction of change (increase or decrease with respect to resting state).

### Identify Sensors susceptible to Music

Music listening proved to invoke network reorganization that was reflected in the detected communities. Hence, we considered the detection of ''mobile'' nodes which systematically participate in the restructuring of functional modules as an important step for the complete description of the music induced self-organization. Following the algorithmic procedure introduced in the corresponding section of methods, we derived a node-indexed profile $N_{sw}$ for each brain rhythm. These profiles, included in Fig.8c, express the tendency of sensors (i.e. of the cortical areas beneath) to change their community membership during music-listening. The nodes of distinctively higher propensity to be influenced by music are the sensors FP1 and P3 participating in the **δ**-rhythm functional network and the sensor PZ participating in the **β$_H$** rhythm network.

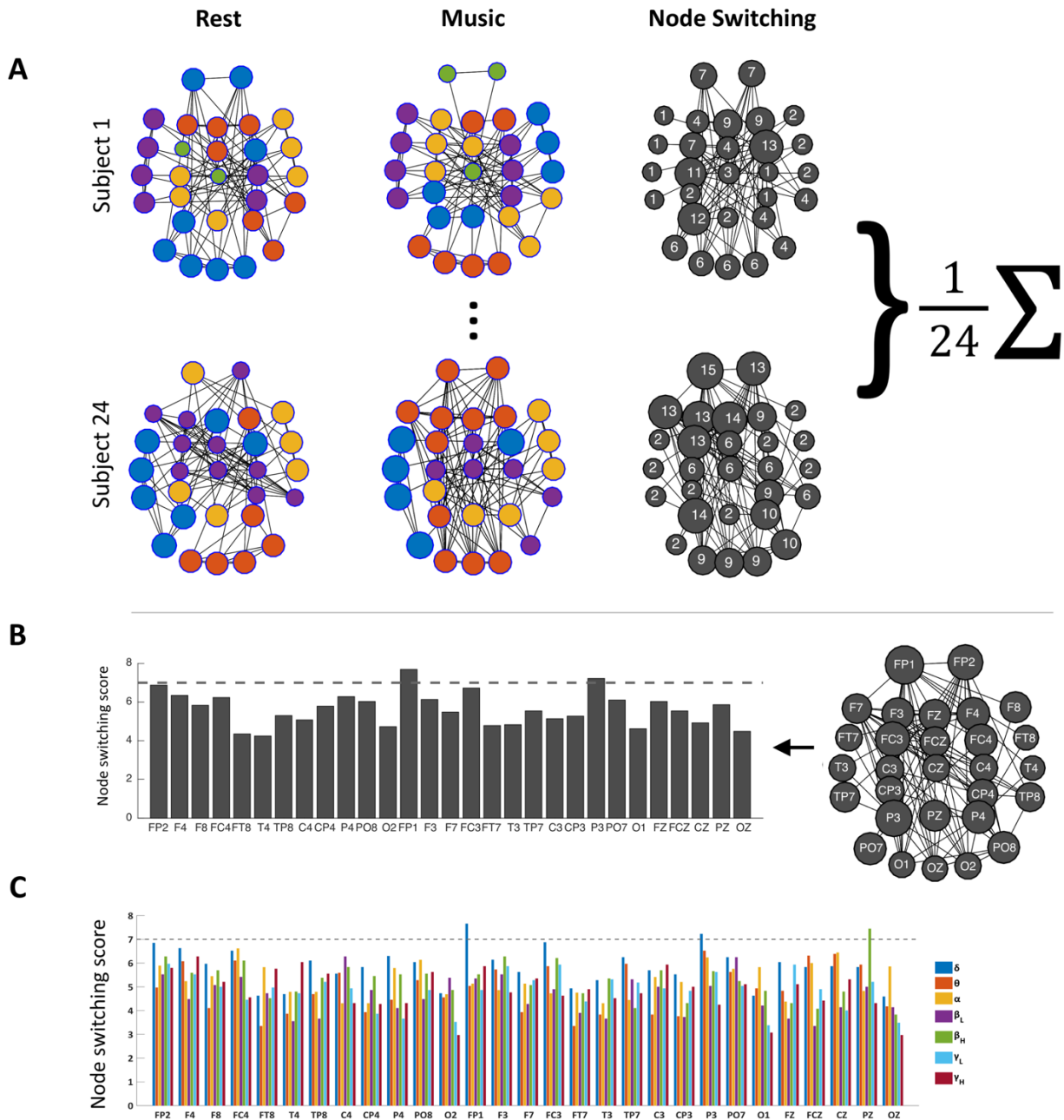

Figure 8: A-B) Exemplifying the algorithm for identifying the switching nodes during music listening using the modular patterns derived from δ-rhythm brain activity. For each subject, the $A^{diff}$ binary connectivity graph (shown in right most column) is constructed over the 29 sensors (based on the comparison between the community patterns detected in resting state and musing listening). By estimating the node degrees of this graph, a subject-specific profile is derived. These profiles are then averaged across subjects to form a rhythm-specific aggregate profile that expresses the propensity of a node to change its grouping during music listening. C) Superposition of the node switching profiles from all brain-rhythms.

## Cross-frequency Coupling in relation to the observed multiplexing

From our analysis, presented so far, there was accumulated evidence about the convergent re-organization between the functional networks related with $\delta$ and $\beta_H$ brain rhythms. This evidence led us to further examine the particular between-rhythm coupling from the perspective of nested oscillations as currently is adopted for explaining cognitive processes [41]. Working independently for each subject and sensor, we first derived estimates of PAC-levels and formed distinct distributions for each recording condition. Next, the distribution of either listening condition was statistically compared with the ''baseline'' distribution corresponding to resting state. Statically significant changes were then detected (p<0.05; FDR corrected) for each listening condition and presented topographically in Figure 9. Interestingly, the adopted procedure resulted in the detection of four (4) ''music-sensitive'' nodes, two of which (FP1, PZ) had been previously detected as nodes of increased music-induced mobility (see Fig.8).

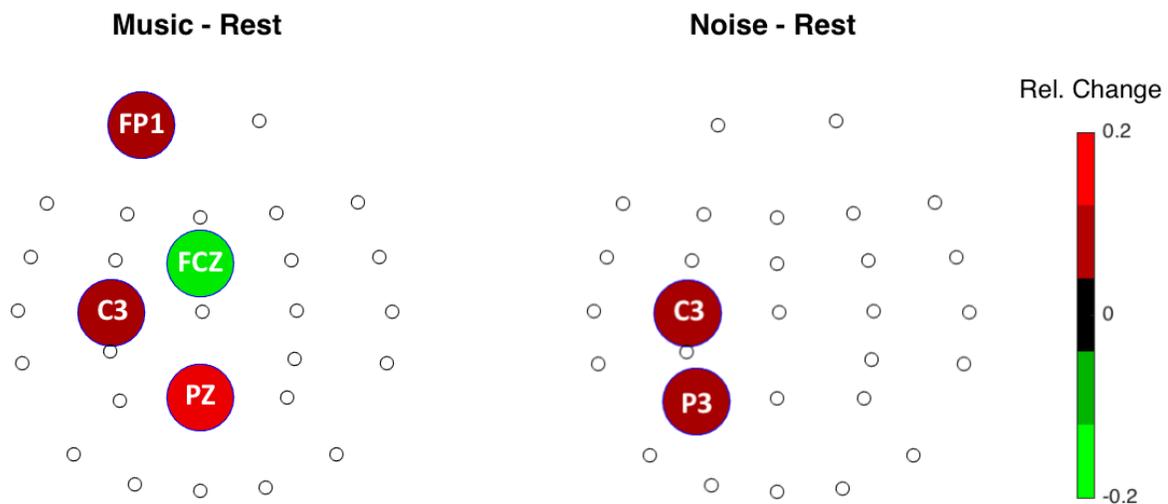

Figure 9. Significant changes in Phase-Amplitude coupling for $\delta => \beta_H$ interaction. PAC-levels were estimated at the level of individual sensors for all subjects and recording conditions. The measurements were then aggregated across subjects, for each recording condition separately, and used to express the relative change. Non-significant changes have been filtered out based on a statistical test (Wilcoxon rank-sum test; p<0.05, FDR-corrected) and here only the important ones are indicated. Left/Right includes the topographical representation of the detected changes in $\delta => \beta_H$ coupling when listening to music/noise. Color indicates both the strength of relative change and the direction of change (increase or decrease with respect to resting state).

# Discussion

The complex networks perspective, as adopted in the study of brain's functional organization, has already deepened our understanding about the human cognition and behavior [34]. Similarly, the approach of multi-scaled complex networks opens new perspectives. It is suggested that the characterization of brain's functional organization should accommodate multiple temporal and spatial scales [7–9,35]. Multilayer networks appear as a powerful concept for encapsulating its inherent (structural and functional) complexity.

We introduce here a novel methodology that combines the derivations from mono-layer network analysis (applied multiple times to individual layers) as a means to detect multi-layer organization. Modular patterns are detected for each brain rhythm (reconstructed network layer) and the corresponding segregations are compared across layers and contrasted between recording conditions. In our algorithmic framework, increased similarity indicates interlayer coordination, network sites that change their membership correspond to interaction locations and PAC serves as the explanatory mechanism behind the observed functional reorganization.

Studying the music-induced changes in patterns of neural synchrony within and across the single-layered networks of the well-known brain rhythms, we provide evidence about the functional inter-layer coherence between $\delta$ and $\beta_H$ rhythms that can be interpreted as increased inter-dependence and hence elevated **multiplexing** as a result of music perception (Fig 7). Moreover, we observe that the noise seems to only affect (less intensely) the reorganization in $\delta$ rhythm but not in $\beta_H$ rhythm. Motivated by these observations, we further examine the nested $\delta \rightarrow \beta_H$ oscillatory coupling, since a similar trend has been reported in recent studies [39]. Interestingly, the two distinct methodological approaches (functional re-organization and nested oscillations) led to commonalities. The hypothesized music-induced cross-frequency interactions revealed two "sensitive" nodes (Fig.9) that were also detected as nodes of high music-induced-*mobility* during the functional reorganization analysis step (Fig.8). Finally, it should be emphasized that conventional spectral analysis did not lead to activation changes that could be attributed to music listening (Fig.1 and supplementary Fig.S1).

At this point, some explanations about specific choices in our study need to be provided. Following is a list of justified choices which implicitly points to the limitations of this study as well: **1)** An eyes-closed experimental paradigm was considered better aligned with the passive character of the music-listening task, avoiding distraction of the subjects and limiting the ocular artifacts in the recorded signal. **2)** The ''superficial'' character of the adopted methodology, with networks being reconstructed in sensor space and serve only as a rough approximation of the underlying cortical networks, was motivated by the exploratory nature of the study and the perspective of finding application in real-life situations, where real-time execution needs to be guaranteed. This particular choice was also reinforced by relevant recent studies[39]. **3)** The selection of phase synchrony was motivated (apart from our previous studies in human cognition) by the results of a MEG study of music perception, in which it was demonstrated that only phase -and not energy- descriptors suffice for the detection of music influences on brain activity [59]. This was also verified by our results (Fig.3 and Fig.S1), since the detected spectral alterations were not music-specific and appeared in noise-listening condition as well. **4)** Phase-synchrony was estimated by means of a simple estimator (PLV), despite the fact that others more sophisticated ones exist in the literature [60]. This particular choice was dictated by a) the fast implementation of PLV metric (compatible with the notion of real-life applications) and b) the fact that it shares the same algorithmic steps with the adopted PAC estimator (compare eq(1) with eq(A5) in appendix). Beyond implementation arguments, we also investigated the possibility that PLV measurements could have been extensively contaminated via volume conduction effects. We provide as supplementary material, the results from the analysis performed so as to grossly estimate the extent of volume conduction in our data (see supplementary Fig.S2). The portion of instantaneous phase differences around zero or ±π (which could be due to volume conduction), was less than 13% for **δ** rhythm and less than 12% for **$β_H$** rhythm. Apart from these indications, we need to underline that throughout this study, results were derived and reported comparatively. Hence, even volume conduction interfered with the true phase-interaction measurements, its effects were expected to be similar in all recording conditions and therefore could not have contributed significantly towards the detected alterations.

Next, we need to place our work in a wider neuroscientific context and discuss our main findings in relation with established knowledge in the neuroscience of music perception. As it is reported in various functional neuroimaging studies, music stimulates the integration of cortical processes with fundamental subcortical rewarding mechanisms in the brain [61]. However, the functional atlas of neural underpinnings in music perception is still largely unclear, as music induces complex and integrating cognitive processing due to its various constituent components like melody, harmony, pitch, tempo and timbre [62]. The assessment of music induced modulations of ongoing brain dynamics is getting increased attention lately, with electroencephalography (EEG) studies reporting changes in the functional organization of neural synchronies during various listening tasks [26–32]. Similarly, our results demonstrate a music-induced functional reorganization based on the analysis of recorded brainwaves patterns. But in direct contrast with previous work, here we also report that this reorganization proceeds in synchrony across the distinct layers of brain rhythms. EEG, compared to other neuroimaging techniques, has the additional advantage (and increasingly gains popularity) in real-world environments due to advances in its portability, credibility and ease of use [63]. Associated with *in situ* neuroimaging capabilities, EEG has already been suggested as a valuable screening facility during music therapy treatment [64] and researchers already discuss the neural underpinnings of music perception within the context of music therapy [24,65–67]. It is noteworthy that in these studies, the role of **β**-band cortical rhythm has been emphasized.

In the field of Magnetoencephalography (MEG), the role of **β**-cortical rhythm has been emphasized and associated with temporal predictions of hierarchical rhythmic patterns in naturalistic music [68]. More intriguingly and in consistency with our findings, auditory and motor cortical areas have been associated with nested **δ** and **β** fluctuations, with the phase of the **δ**-rhythm modulating the power of the **β**-rhythm (see a review in [69]). This particular type of cross-frequency coupling between **δ** and **β** rhythms has already been suggested to control the precision of neural oscillations for the temporal predictions of auditory targets [39] and found to operate in cortical areas with a similar localization with our PAC-related results (Fig.9). Hence, considering the empirical evidence of functional interdependence between **δ** and **β**$_H$ rhythms (Figures 5 and 8), our results reinforce previous evidence [41] which suggests that PAC could constitute a mechanism for the multiplex organization in cortical networks.

# Conclusion

The present study introduces a novel, elegant and multiplex-aware approach that for the first time encapsulates the inter-layer dependencies of cortical neural oscillations in the context of brain oscillatory organization in music perception. Our computational approach can be easily implemented in conjunction with alternative measures of functional connectivity and used in real-life applications such as music recommendation systems [23] and intangible artistic interactive interfaces [70].

# Appendix

## Appendix 1: Functional segregation using the dominant-sets algorithm

The algorithm searches for the most-cohesive group of vertices, given an undirected weighted matrix **W** of a graph; then, working in a subtractive fashion, it is iteratively applied until it finally yields the effective clustering of pairwise-relational data residing in the original graph; a detailed mathematical formulation may be found in a previous study [45]. One of its main characteristics is its compact and elegant formulation, taking the form of deriving the N-dimensional vector **x** that maximizes the following objective function of **Cohesiveness** (i.e., the set-compactness):

max F(x) = **x**$^T$ **W x**,

$$\mathbf{x} \in \Delta, \Delta^n = \left\{ x \in R^n : x_i \geq 0 \; \forall \; i \; \cup \; \sum_{i=1}^{n} x_i = 1 \right\} \quad (A1)$$

The algorithmic procedure is denoted as:

$$\{\mathbf{x}, F(\mathbf{x})\} = \text{Dominant Set}(A) \quad (A2)$$

The vector **x** lists the memberships for all nodes in the graph and can be used to identify the exact list of nodes participating in the dominant-set (by locating the non- zero elements). The second output F(**x**) is the particular value of objective function that measures the cohesiveness of the detected dominant-set. Following a fixed number of iterations, the set of indices corresponding to the nonzero components of **x** is computed, resulting the set of nodes participating in the dominant graph-component. The full partition of **W** into disjoint sets of nodes is finally accomplished by repeating the following three steps: (i) finding the current dominant set, (ii) removing the vertices in that cluster, and (iii) iterating again on the rest of nodes.

## Appendix 2: The Variation of Information (VI) measure

VI is defined by the function:

$$\text{VI}(\mathbf{c}, \mathbf{c}') = \big(H(\mathbf{c}) - I(\mathbf{c}, \mathbf{c}')\big) + \big(H(\mathbf{c}') - I(\mathbf{c}, \mathbf{c}')\big) \quad (A3)$$

where H(**c**) denotes the entropy associated with cluster profile **c** and is defined by the function:

$$H(\mathbf{c}) = -\sum_{k=1}^{K} P(k) \log P(k) \quad (A4)$$

and $P(k) = n_k/n$ is the probability of an item belonging to cluster $c_k$. $I(\mathbf{c}, \mathbf{c}')$ denotes the mutual information between the two cluster profiles $\mathbf{c}$ and $\mathbf{c}'$. A Matlab® implementation of VI can be found in the Brain Connectivity Toolbox[4], under the function *partition_distance*.

## Appendix 3: Phase-Amplitude Coupling based on phase coherence

Described in a more generic setting, let y(t), t = 1, 2, ..., T is the recorded is the recorded single-sensor signal at hand. Based on filtered versions of this signal, cross-frequency interactions will be sought complying with a form in which the phase of low-frequency oscillations modulates the amplitude of high-frequency oscillations. Using narrowband filtering, the two signals $y_L(t)$ (low-frequency) and $y_H(t)$ (high-frequency) are first formed and, then, their complex analytic representations $z_L(t)$ and $z_H(t)$ are derived by means of Hilbert transform (HT[.]).

$$z_L(t) = HT[y_L(t)] = |y_L(t)|e^{i\,\varphi_L(t)} = A_L(t)e^{i\,\varphi_L(t)}$$

$$z_H(t) = HT[y_H(t)] = |y_H(t)|e^{i\,\varphi_H(t)} = A_H(t)e^{i\,\varphi_H(t)}$$

In this way the amplitude and phase dynamics, captured respectively by the envelope A(t) and instantaneous phase φ(t) signal, can be treated independently. Next, the envelope of the higher-frequency oscillations $A_H(t)$ is bandpass-filtered within the range of low-frequency oscillations and the resulting signal undergoes an additional step of Hilbert transform so as to isolate its phase- dynamics component $\varphi'(t)$,

$$z'_H(t) = HT[A_{L,H}(t)] = |z'(t)|\,e^{i\,\varphi'_H(t)} = |z'(t)|\,e^{i\,\varphi_{L\to H}(t)}$$

that reflects the modulation of HF-oscillations amplitude by the phase of the LF-oscillations. The corresponding time series will be used to estimate PAC, by means of phase-locking (or synchronization index) technique.

$$PLV_{L\to H} = \left|\frac{1}{T}\sum_{t=1}^{T} e^{i\,(\varphi_L(t) - \varphi'_H(t))}\right| \quad (A5)$$

Phase-locking value PLV ranges between 0 and 1, with higher values indicating stronger PAC interactions (i.e., higher comodulations).

---

[4] https://sites.google.com/site/bctnet/